\documentclass[11pt,a4paper]{article}
\usepackage{jcappub}
\usepackage{subfig}
\usepackage{bm}

\newcommand{\Mpl}{M_{\mathrm{P}}}

\makeatother

\begin{document}

\title{Bounds from ISW-Galaxy cross-correlations on generalized covariant Galileon models}

\author[a,b]{Francesco Giacomello}
\author[c]{Antonio De Felice}
\author[b,d,e]{Stefano Ansoldi}
\affiliation[a]{Dipartimento di Fisica, Universit\`{a} degli Studi di Trieste, Italy}
\affiliation[b]{Department of Physics, Kyoto University, Japan}
\affiliation[c]{Center for Gravitational Physics, Yukawa Institute for Theoretical Physics, Kyoto University, 606-8502, Kyoto, Japan}
\affiliation[d]{Istituto Nazionale di Fisica Nucleare (INFN), Italy}
\affiliation[e]{Department of Mathematics, Computer Science, and Physics, Udine University, Italy}

\emailAdd{francesco.giacomello1@gmail.com}
\emailAdd{antonio.defelice@yukawa.kyoto-u.ac.jp}
\emailAdd{ansoldi@fulbrightmail.org}

\abstract{Several modified cosmological models exist, which also try to address the tensions between data and predictions of the $\Lambda$-CDM model. Galileon models are particular scalar tensor theories that represent one such possibilities. While it is commonly understood that there may be inconsistencies between predictions of some Galileon models and observations, in particular concerning ISW-Galaxy cross-correlations, there is no proof yet that these models are completely ruled out. Indeed, by using a specific background (known as the \emph{the tracker solution}) in the generalized covariant Galileon theory, we show that, after imposing all standard theoretical stability constraints, it is still possible to identify a region in the parameter space of the model that allows for positive ISW-Galaxy cross-correlation. By a physical interpretation in terms of a chi-square analysis, we confirm the expectation that in this \emph{viable} region the predictions of generalized covariant Galileon theory on the tracker solution background have higher likelihood when they approach the physics of the $\Lambda$-CDM model.}


\maketitle

\section{Introduction}

The late time Universe accelerated expansion is supported by observational evidence, such as the Cosmic Microwave Background~(CMB)~\cite{WMAP1999}, baryon acoustic oscillations~\cite{SDSS2005}, and supernovae Ia~\cite{9805201}. These evidences witness an astonishing increase in precision of observations in cosmology, and allow for an unprecedented opportunity for quantitative tests on theoretical models.

The simplest way to model the so called Dark Energy (DE), responsible for the observed cosmic acceleration, is to postulate the existence of a cosmological constant. However, this explanation is not free from problems, above all the fact that the DE scale is incredibly smaller than the typical particle physics vacuum energy~\cite{Weinberg89}. This is one of the reasons why a lot of alternative models have been proposed, providing different models of DE~\cite{DEmodels,1753}. Generally speaking, these models can be classified into two kinds: modified matter models, such as quintessence~\cite{quintessence} or k-essence~\cite{kessence}, and modified gravity models. In this latter class of theories we find, among others, $f(R)$ theories~\cite{fR1, fR2}, $f(R,\mathcal{G})$ theories~\cite{fRG}, Brans-Dicke theories~\cite{BD}, Dvali-Gabadadze-Porrati (DGP) braneworld models~\cite{DGP}, and, finally, Galileon gravity~\cite{gal1,gal2}.

In this paper, we will focus on a particular generalized Galileon theory. Galileon theories are a particular subset of Horndeski theories~\cite{Horndeski}, which include all scalar-tensor theories having second order field equations in a four dimensional space-time. For many Galileon theories, such as for example the quartic and quintic Galileon models, the predicted propagation speed of  the tensor perturbation modes is such that $c_{T}^{2}\neq1$. However, the latest multimessenger astrophysical observations rule out such models~\cite{ligo}.

Another important observable, which is used to constrain DE models, is the integrated Sachs-Wolfe~(ISW) effect, and, in particular (see section~\ref{sectionISW}), the cross correlation between the ISW signal and the galaxy distribution. The standard $\Lambda$-CDM model predicts the sign of this cross correlation to be positive, in agreement with current observations. On the other hand, it has been pointed out that the cubic Galileon model (and also the quartic and quintic models for most of the parameter space), generally, predicts a negative cross correlation~\cite{0485,02263,4641}.

Here, we will show that what happens for the cubic Galileon model does not rule out the viability of other Horndeski theories with $c_T^2=1$. In particular, we will show that there is a region in the parameter space of the solution to the generalized Galileon model introduced in~\cite{3878} that satisfies $c_{T}^{2}=1$, and allows for positive ISW-Galaxy cross correlation.

The paper is organized as follows: in section~\ref{sec:model} we review the Galileon model that we consider in our analysis. Stability conditions and other physical conditions on the model are then discussed in section~\ref{stability conditions}. Basic ideas about ISW-Galaxy cross correlation can be found in section~\ref{sectionISW}, with the goal of clarifying technical details of our approach. After discussing some relevant approximations in section~\ref{sectionapprox}, in section~\ref{sectionresults} we present the results of our numerical analysis and provide their physical interpretation. Finally, we conclude with some general remarks in section~\ref{sectionconclusion}.

\section{\label{sec:model}The model}

We consider a particular form of the general Horndeski Lagrangian~\cite{Horndeski} with two additional matter components, $\mathcal{L} _{\mathrm{m}}$ and $\mathcal{L} _{\mathrm{r}}$, in the form of barotropic perfect fluids with energy densities $\rho _{\mathrm{m}}$ and $\rho _{\mathrm{r}}$ that represent pressure-less dust and radiation, respectively. Then the Lagrangian can be written as
\begin{equation}
	\mathcal{L}
	=
	K (\phi , X)
	-
	G _{3} (\phi , X) \Box \phi
	+
	\frac{M _{\mathrm{P}} ^{2}}{2}
	R
	+
	\mathcal{L} _{\mathrm{m}}
	+
	\mathcal{L} _{\mathrm{r}}
	\,,
\label{eq:lag}
\end{equation}
where $X = ( \nabla \phi ) ^{2}$, for a generic function $A(\phi,X)$ we adopt the notation $A_{,X}\equiv\partial A/\partial X$, and $A_{,\phi}\equiv\partial A/\partial \phi$, while $M _{\mathrm{P}}$ is the Planck mass.

Additionally, we restrict our attention to the subclass of models for which
\begin{eqnarray}
	K & = & -c_{2}M_{2}^{4-4p}X^{p}\,,\\
	G_{3} & = & c_{3}M_{3}^{1-4p_{3}}X^{p_{3}}\,,
\end{eqnarray}
where $p _{3}$ is just a temporarily convenient replacement for a variable $q$ that will turn out to be more useful in the following discussion:
\begin{equation}
	p _{3} = p + q - \frac{1}{2}\,.
\end{equation}
Moreover, $M _{2}$ and $M _{3}$ are two other mass scales, and $c _{2}$ and $c _{3}$ dimensionless constants. This sub-class of models has the property that the propagation speed of the tensor modes is $c_T^2=1$. This property does not depend on the particular fine-tuned FLRW solution, nor on the choice of the background. In other words, independently of the dynamics and on the background, the tensor modes always propagate at the speed of light.

If we consider a flat FLRW background, and denote by $a(t)$ the conformal factor as a function of the local proper time $t$, Friedmann equation takes the form
\begin{equation}
3H^{2}M_{\mathrm{P}}^{2}=\rho_{\rm DE}+\rho_{\mathrm{m}}+\rho_{\mathrm{r}}\,,
\end{equation}
where
\begin{equation}
\rho_{\rm DE}\equiv 2XK_{,X}-K+6H\dot{\phi}XG_{3,X} \,
\end{equation}
is the Dark Energy component modelled by the scalar field $\phi$. Friedmann equation admits a solution~\cite{3878}, known as the ``tracker'' solution, that satisfies
\begin{equation}
H\dot{\phi}^{2q}={\rm constant}\,,\label{eq:defTrack}
\end{equation}
where, $H = d \mathcal{N} / (dt)$ is the Hubble parameter, and $\mathcal{N}$ the number of e-folds $\mathcal{N}=\ln(a/a_{0})$ (with standard notation, $a _{0}$ is the scale factor at the present epoch). A characteristic property of this solution is the fact that $\dot{\phi}$ becomes constant for large $t$. This means that $H$ becomes constant as well in the same limit, and spacetime approaches a late-time de~Sitter (dS) background. This justifies the following notation for the constant that appears in~(\ref{eq:defTrack}), $(H_{{\rm dS}}\Mpl x_{{\rm dS}})^{2q}$, which provides an implicit definition for $H _{\mathrm{dS}}$, the asymptotic value for $H$, and for $\dot{\phi} _{\mathrm{dS}}$, the asymptotic value for $\dot{\phi}$. Additionally, $x = \dot{\phi} / (H M _{\mathrm{P}})$, is a dimensionless replacement for $\dot{\phi}$, that can be conveniently used in the computations. Also the mass scales ($M _{2}$ and $M _{3}$) and the dimensionless constants ($c _{2}$ and $c _{3}$) can be rewritten in terms of the quantities defined just above: 
\begin{eqnarray}
M_{2} & = & \sqrt{H_{{\rm dS}}\,\Mpl}\,,\\
M_{3} & = & \left(\frac{\Mpl^{1-2p_{3}}}{H_{{\rm dS}}^{2p_{3}}}\right)^{1/(1-4p_{3})}\,,\\
c_{2} & = & 3\left(\frac{2}{x_{{\rm dS}}^2}\right)^{p},\\
c_{3} & = & \frac{\sqrt{2}p}{2p+2q-1}\left(\frac{2}{x_{{\rm dS}}^2}\right)^{p+q}\,.
\end{eqnarray}
We conclude with a final change of dynamical variables, mostly motivated by convenience. In the new variables
\begin{align}
r_{1} & \equiv \Bigl(\frac{x_{\rm dS}}{x}\Bigr)^{2q}\biggl(\frac{H_{\rm dS}}{H}\biggr)^{1+2q}\,,\\
r_{2} & \equiv \Biggl[\biggl(\frac{x}{x_{\rm dS}}\biggr)^{\! 2}\frac{1}{r_{1}^{3}} \Biggr]^{\frac{p+2q}{1+2q}}
=\left(\frac{\dot{\phi}}{x_{{\rm dS}}\Mpl H}\right)^{2(p+2q)/(1+2q)}=\left(\frac{\dot{\phi}}{x_{{\rm dS}}\Mpl H}\right)^{4q(1+s)/(1+2q)}\,,\\
\Omega_{\mathrm{r}} & \equiv \frac{\rho_{\mathrm{r}}}{3H^{2}\Mpl^{2}}\,,
\quad
\Omega_{\mathrm{m}} \equiv \frac{\rho_{\mathrm{m}}}{3H^{2}\Mpl^{2}}\,,
\quad
\Omega_{\mathrm{DE}} \equiv \frac{\rho_{\mathrm{DE}}}{3H^{2}\Mpl^{2}}\,,
\end{align}
the late time, dS fixed point corresponds to $(r_{1},r_{2},\Omega_{\mathrm{r}})=(1,1,0)$, and the tracker solution corresponds to $r_1=1$ during the whole evolution. Different equivalent expressions for $r _{2}$ are given above for convenience, and we also introduce the parameter
\begin{equation}
s = \frac{p}{2 q}\,,
\end{equation}
as it will be sometimes convenient to use the pair $(s,p)$, while the pair $(s,q)$ will be preferred other times. Given the simple form of $r _{1}$ on the tracker solution, we can concentrate our attention on the equation for $r _{2}$, which allows us to find the behaviour of the scalar field as well. This is
\begin{equation}
r'_{2}=\Omega'_{{\rm DE}}=\frac{3\left(s+1\right)r_{2}\left(1+\Omega_{\mathrm{r}}/3-r_{2}\right)}{1+s\,r_{2}}\,, \label{eq:r2ev}
\end{equation}
where a prime denotes a derivative with respect to $\mathcal{N}$. Since we also have $\Omega_{\mathrm{m}}+\Omega_{\mathrm{r}}+\Omega_{{\rm DE}}=1$, then $0\leq r_{2}\leq1$. Although the previous
equation can not be integrated, in general, in terms of elementary functions, we find that the dS solution $r_{2}=1$ is a fixed point. Moreover, during dust domination, the DE density is negligible ($r_{2}\ll1$) and $\Omega_{\mathrm{r}}\approx0$, therefore we find
\begin{equation}
r'_{2}\approx3\left(s+1\right)r_{2}\,,
\end{equation}
which has approximate solution (we set the value of the scale factor at the current epoch, $a _{0}$, equal to unity)
\begin{equation}
r_{2}\propto e^{3(s+1)\mathcal{N}}=a^{3(s+1)}\,.\label{eq:sol_r2_early_times}
\end{equation}
From the above we see that, in order to have a Dark Energy component, which grows in time (at least during matter domination), a sufficient condition is
\begin{equation}
s+1>0\,.
\label{eq:conzer}
\end{equation}
We quote the following relations that are valid on the tracker solution in addition to~(\ref{eq:r2ev}), and that can be useful later on:
\begin{eqnarray}
H & = & H_{{\rm dS}}r_{2}^{-1/(2s+2)}\,,\label{eq:Heq}\\ 
\dot{H} & = & \frac{dH}{dt}=-\frac{H_{{\rm dS}}^{2}\,[(3+\Omega_{\mathrm{r}})\,r_{2}^{-1/(s+1)}-3r_{2}^{s/(s+1)}]}{2\,(1+r_{2}s)}\,,\label{eq:Hdoteq}\\
\ddot{\phi} & = & -\frac{\dot{H}\dot{\phi}}{2qH}\,,\\
\Omega_{\mathrm{r}}'&=&-{\frac {\Omega_{\mathrm{r}} \left[1-\Omega_{\mathrm{r}} +r_2\,(3+4s) \right] }{sr_{{2}}+1}}\,.\label{eq:Omrev}
\end{eqnarray}
From this last set of equations we can immediately deduce a nice property of the model\footnote{\label{pag:profoo}As a side comment, we would like to point out that these same background equations of motion are shared by a completely different model, namely the generalized Proca theories \cite{ProcaTheories}. However, in the model discussed here, the background is achieved only on the attractor solution, whereas in the generalized Proca theory it is a general cosmological solution. Furthermore, the models are essentially different at the level of the dynamics of cosmological linear perturbations. Therefore, what we will find later about the ISW-Galaxy cross-correlation will not hold in general for the Proca models.}: the evolution of the background, given by Eqs.\ (\ref{eq:r2ev}) and (\ref{eq:Omrev}), only depends on the parameter~$s$.

\section{Stability conditions} \label{stability conditions}

In this section, we discuss stability of the flat FLRW background for the action coming from~(\ref{eq:lag}), i.e. we include two perfect fluids~\cite{3878} (for example by implementing the Schutz-Sorkin action). After eliminating all the auxiliary fields, it is possible to write down the reduced action in the following form (we provide a formal discussion here, details can be found in the literature)
\begin{equation}
S^{(2)}=\int d^{4}x \,a^3\Bigl[\mathcal{A}_{ij}\mathcal{\dot{Q}}_{i}\mathcal{\dot{Q}}_{j}-\mathcal{C}_{ij}(\partial \mathcal{Q}_{i})(\partial \mathcal{Q}_{j})+\dots{}\Bigr]\,.
\end{equation}
In the above formula $\mathcal{Q}_{i}$, $i = 1 , 2, 3$, stands for the density perturbations of the matter fields (the scalar field, and the two components fluid), and we are working in the flat gauge on the background discussed in the previous section.

First, the absence of ghost degrees of freedom, i.e. degrees of freedom that would result in an Hamiltonian  unbounded from below~\cite{1981}, can be studied by diagonalizing the kinetic matrix  $\mathcal{A}$. From this, we find one normalized non-trivial condition
\begin{equation}
Q_{S}\equiv\frac{12(1+s r_2)sq^{2}r_{2}^{(2qs-1)/[2q(s+1)]}}{x_{\rm dS}^{2}(1-2qsr_{2})^{2}} >0\,.\label{eq:qs}
\end{equation}
At early times the Dark Energy density will be negligible, i.e.\ $r_{2}\to0$ (see the description of the background in the previous section; here we are assuming~(\ref{eq:conzer}), but we will see just below that we can actually find a more restrictive condition). In this case, in order to avoid strong coupling problems at early time, we also require the exponent of $r_{2}$ in~(\ref{eq:qs}) to be non-positive
\begin{equation}
\frac{2qs-1}{2q(s+1)}\leq0\,.\label{eq:sc_earlytimes}
\end{equation}
At the same time, when we consider the late time dS limit, $r_{2}\to 1^{{-}}$ and $\Omega_{\mathrm{r}}\to0^{{+}}$, $Q_S$ stays positive, but, in order to avoid strong coupling, we now need
\begin{equation}\,
s(q/x_{\rm dS})^{2}>0\,.\label{eq:no_ghost}
\end{equation}
Computationally, this will be implemented as $s(q/x_{\rm dS})^{2} > \epsilon$, where $\epsilon$ is some small quantity ($\epsilon=10^{-2}$ turns out to be sufficient for our goal). Once we fix both the parameters $q$ and $s$, the condition above will, in turn, limit the allowed values for $x_{\rm dS}$.

Remembering now~(\ref{eq:conzer}), and considering that~(\ref{eq:no_ghost}) already tells us that $s>0$, we see how~(\ref{eq:sc_earlytimes}) has no solution for $q<0$. So far, we can summarize all these constraints as
\begin{eqnarray}
1-2qs & \geq & 0\,,\label{eq:cond1} \\
q & > & 0\,,\\
s & > & 0\,.
\end{eqnarray}

Now we proceed to exclude Laplacian instabilities, by imposing the square of the scalar mode propagation speed to be non-negative. A violation of this condition would lead to an exponential growth of the perturbations in a time-interval much smaller than the Hubble time~\cite{1981}. The required speed of propagation, $c_{s}$, is given by the discriminant equation
\begin{equation}
\det(c_{s}^{2}\mathcal{A}-a^{2}\mathcal{C})=0\,,
\end{equation}
which, on our background, leads to the non trivial condition
\[
c_s^2={\frac {\left[ 3-8\,s{q}^{2}-
 \left( 8\,s+6 \right) q \right] r_{{2}}-8\,r_2^{2}{q}^{2}{s}^{2}+ \left[\left( 4\,s+2
 \right) q-1 \right] \Omega_{{r}}-3+ \left( 12\,s+10 \right) q}{24 \left( 
sr_{{2}}+1 \right) ^{2}{q}^{2}}}\geq 0\,,
\]
where $0\leq\Omega_{\mathrm{r}}\leq1$ parametrizes the radiation component.

Considering again the early time limit, $r_{2}\to0$, in a Radiation Domination (RD) scenario (for which $\Omega_{r}=1$ and $r_2=0$), we obtain
\begin{equation}
c_{s,{\rm RD}}^{2}=\frac{q(4s+3)-1}{6q^{2}}\geq0\,.
\end{equation}
The same reasoning, during Matter (dust) Domination (MD), $\Omega_{\mathrm{r}}=0$ and $r_2=0$, results in
\begin{equation}
c_{s,{\rm MD}}^{2}=\frac{2q\,(6s+5)-3}{24q^{2}}\geq0\,.
\end{equation}
Finally, by taking the limit of a dS universe, $\Omega_{\mathrm{r}}=0$ and $r_2=1$, one finds
\begin{equation}
c_{s,{\rm dS}}^{2}=\frac{1-2qs}{6q\,(1+s)}\geq0\,,
\end{equation}
which does not add any constraint since it is now equivalent to the inequality~(\ref{eq:sc_earlytimes}).

One can perform a similar analysis on tensor modes, but, in fact, they do not result in additional constraints, since $c_T^2=1$ (this was, indeed, one of the reasons behind the choice of the original model; moreover, vector modes also do not add additional constraints), and
$Q_{T}=\Mpl^{2} / 4$.

Putting together all the previous conditions we find the region of parameter space, which is accessible on theoretical grounds, to be
\begin{equation}
s>0\qquad{\rm and}\qquad\frac{1}{3+4s}\leq q\leq\frac{1}{2s}\,.\label{eq:qspace}
\end{equation}
Since $s$ can be, in principle, arbitrary small, the upper bound for $q$ can, in general, be arbitrarily large, and it would be impractical to work in a region that is not bounded. On the other hand, if $q$ tends to saturate the upper bound above, even for small values of $s$, then $2qs\simeq1$, and, as we will see later on, in this case, $G_{\rm eff}/G_N\to\infty$ on the dS background. In turn, such a limit tends to result in a bad fit to the ISW-Galaxy correlation data. Furthermore, as we have already mentioned in the footnote on page~\pageref{pag:profoo}, the background for this Horndeski theory is equal to the cosmological background of the Proca theory. For the Proca case, it has been found~\cite{ProcaTheories} (for a less recent background constraint analysis see also~\cite{CosmTsuji} and, for covariant Galileons, \cite{Peirone}), that the best fit to the background cosmological data leads to $s\geq s_{\rm min }>0$. The reason for such a behaviour is the fact that a strictly positive value for $s$ mitigates the tension between late-time measurements of $H_0$ and estimations by Planck, when using the time evolution of the background equations of motion. In this sense, we expect an upper limit for $q$ to appear naturally, once we implement the background constraints on the parameter~$s$.

\section{ISW-Galaxy cross correlation} \label{sectionISW}

The Integrated Sachs-Wolfe (ISW) effect is responsible for the creation of CMB anisotropies affecting photons which encounter time varying gravitational potentials during their journey towards us. Large scale potential wells (or hills) can vary significantly during the period of time taken by a photon to travel through them, because of the accelerated expansion of the universe. Thus if, for example, a photon enters a potential well which is getting shallower, then the photon itself will undergo a net energy gain. It is then intuitive that the ISW effect can provide useful indications about the nature of DE.  However, the ISW signal by itself is quite weaker than other CMB anisotropies: this is usually addressed by computing the cross correlation between the ISW effect and some large scale structure tracker, which, effectively, enhances observables for this effect. For example, it is reasonable to expect a cross correlation with the matter distribution, since the time evolution of the gravitational potential is related to the Dark Matter distribution by the Poisson equation~\cite{01720}. Indeed, we will compute the cross correlation with the galaxy density distribution (see Eq.~(\ref{eq:gdens}) later). Notice that, in principle, the ISW effect can be generated by scalar, vector and tensor perturbation modes; however, the significant contribution at linear level comes from the scalar modes~\cite{5102}.

In order to introduce the ISW observable, we will employ the field $\psi_{\rm ISW}\equiv\Psi+\Phi$ (we follow the definition of all these quantities used in~\cite{3878}), so that
\begin{equation}
 \frac{\Delta T_{\rm ISW}(\eta,\boldsymbol{\hat{\theta}})}{T}=\lim_{\eta \to \eta_{0}}\int_{\eta _{i}}^{\eta}d\eta \frac{\partial \psi _{\rm ISW}}{\partial \eta}=-\lim_{z\to0}\int_{z}^{z_{i}}dz\frac{\partial \psi_{\rm ISW}}{\partial z}\,,
\end{equation}
where $\eta_{i}$ and $z_{i}$ correspond, in our case, to some chosen values of conformal time and redshift in the deep matter era.

Another important quantity in our analysis is the density of galaxies. Here we make the assumption that in the matter perturbation $\delta(\bm{k},a)$ we can separate the time, $a$, and mode, $k$, dependences. This can be done~\cite{01790} when the equation of motion for $\delta$ depends only mildly on $k$. In this case we can write \footnote{In what follows, we will use interchangeably dependences from $a$, $z$, ${\mathcal{N}}$, \dots{} without changing the name of the functions.} $\delta(\bm{k},a)\propto D(a)$, or $\delta(\bm{k},z)=D(a)/D_0\,\delta(\bm{k},z=0)$, from which one finds
\begin{equation}
\frac{d\delta}{d\mathcal{N}}=\frac{dD}{d\mathcal{N}}\frac{\delta(\bm{k},0)}{D_{0}}\,.
\end{equation}
Finally we can introduce the observed projected galaxy overdensity $g$,
\begin{equation}
g=\int_{0}^{z_{r}}dzb(z)\phi(z)\delta(z,\chi,\hat{n})\,,\label{eq:gdens}
\end{equation}
where $\chi$ is the comoving radial coordinate, $b(z)$ is the galaxy bias and $\phi(z)$ is a window function of the form \cite{4380}
\begin{equation}
\phi(z)=\frac{\beta}{\Gamma[(m+1)/\beta]}\biggl(\frac{z}{z_{0}}\biggr)^{m}e^{-(z/z_{0})^{\beta}}\,,
\end{equation}
where $\beta$, $m$, and $z_{0}$ are parameters fixed by the observation, and the normalization is set so that
\begin{equation}
\int_{0}^{\infty}\phi(z)dz=1\,.
\end{equation}

Notice that in (\ref{eq:gdens}) the galaxy bias is considered, in general, as a redshift dependent quantity; however, we will select only those ISW observations for which the window function is very peaked around a particular redshift. In this way we will be able to neglect the $z$-dependence of the bias, and our results will, thus, be independent from variations of the bias with time.

In order to find the ISW-Galaxy cross correlation one has to evaluate the two-point function
\begin{equation}
C_{l}^{\rm GI}=\langle a_{lm}^{\rm ISW}a_{lm}^{{\rm G}*}\rangle
\,,
\end{equation} 
where an expansion with spherical harmonics has been implemented on both the ISW and galaxy over-density integral. Applying standard results we find
\begin{eqnarray}
C_{l}^{\rm GI} & = &\left\langle a_{lm}^{{\rm ISW}}a_{lm}^{{\rm G}*}\right\rangle
\nonumber \\
& = & 
\frac{2}{\pi D_{0}^{2}}\int_{k_{m}}^{k_{M}}k^{2}dk\,P(k)\int_{\mathcal{N}_{0}}^{\mathcal{N}_{i}}d\mathcal{N}_{1}\,j_{l}(k\chi_{1})Z_{{\rm ISW}}(\mathcal{N}_{1})
\int_{\mathcal{N}_{0}}^{\mathcal{N}_{i}}d\mathcal{N}_{2}e^{-\mathcal{N}_{2}}\,\phi(\mathcal{N}_{2})\,b_{s}\,D(\mathcal{N}_{2})\,j_{l}(k\chi_{2})\,,\label{eq:ClGI}
\nonumber
\end{eqnarray}
where the variable $Z_{\rm ISW}$ is defined in terms of the $\psi_{\rm ISW}$ field as
\begin{equation}
\frac{\partial \psi_{\rm ISW}}{\partial\mathcal{N}}=Z_{\rm ISW}\frac{\delta(0,\bm{k})}{D_{0}}\,,
\end{equation}
and we have introduced the power spectrum $P(k)$, defined as
\begin{equation}
\langle X(\bm{k}_{1})X(\bm{k}_{2})^{*}\rangle=(2\pi)^{3}P(k_{1})\,\delta^{(3)}_{\rm Dirac}(\bm{k}_{1}-\bm{k}_{2})\,.
\end{equation}
The expression of $P(k)$ in terms of the matter transfer function $T_{m}(k)$ is given by
\begin{equation}
P(k)=2\pi^{2}\,\delta_{H}^{2}\,[T_{m}(k)]^{2}\,\left(\frac{k}{H_{0}}\right)^{n_{s}}H_{0}^{-3}\,, \label{eq:powerspectrum}
\end{equation}
where $\delta_{H}^2$ is the power spectrum normalization, and $n_{s}$ is the spectral index. Along the trajectory of a photon
\begin{equation}
\chi=\int_{\eta}^{\eta_{0}}d\eta=\int_{0}^{z}\frac{dz}{H}\,,
\end{equation}
so that we can conveniently define the dimensionless quantity
\begin{equation}
\bar{\chi}=H_{0}\chi=\int_{0}^{z}\frac{H_{0}}{H}dz\,,
\end{equation}
which obeys the equation 
\begin{equation}
\frac{d\bar{\chi}}{dz}=\frac{H_{0}}{H}=\Bigl(\frac{r_{2}}{r_{2,0}}\Bigr)^{1/[2(s+1)]}\,.
\end{equation}
The last equality comes from the background evolution of $H$, given by (\ref{eq:Heq}), and $r_{2,0}=\Omega_{\rm DE,0}$. In terms of the e-fold variable $\mathcal{N}$, $\bar{\chi}$ evolves as
\begin{equation}
\frac{d\bar{\chi}}{d\mathcal{N}}=-e^{-\mathcal{N}}\biggl(\frac{r_{2}}{r_{2,0}}\biggr)^{1/[2(s+1)]}=-\biggl(\frac{\rho_{m}}{\rho_{m0}}\biggr)\biggl(\frac{r_{2}}{r_{2,0}}\biggr)^{1/[2(s+1)]}\,, \label{eq:chibarev}
\end{equation}
where we used $\rho_{m}=\rho_{m0}a^{-3}=\rho_{m0}e^{-3\mathcal{N}}$ . When neglecting radiation at late times, $\Omega_{\mathrm{m}} \approx 1-r_{2}$, and we can rewrite Eq.\ (\ref{eq:chibarev}) as 
\begin{equation}
\frac{d\bar{\chi}}{d\mathcal{N}}=-\biggl(\frac{1-r_{2}}{1-r_{2,0}}\biggr)^{1/3}\biggl(\frac{r_{2}}{r_{2,0}}\biggr)^{1/[6(s+1)]}.
\end{equation}
This differential equation satisfies the initial condition $\bar{\chi}(\mathcal{N}=0)=0$, and should be solved together with~(\ref{eq:r2ev}), the background equation of motion, which, under this approximation, reads
\begin{equation}
r'_{2}  \approx  \frac{3\left(s+1\right)r_{2}\left(1-r_{2}\right)}{1+s\,r_{2}}\,;
\end{equation}
the corresponding initial condition is $r_{2}(0)=1-\Omega_{\mathrm{m}0}$. Finally, in a completely analogous way, we can express the galaxy autocorrelation as
\begin{eqnarray}
C_{l}^{\rm GG}
& = &
\left\langle a_{lm}^{\rm G}a_{lm}^{\rm G*}\right\rangle   
\nonumber \\
& = &
\frac{2}{\pi D_{0}^{2}}\int_{k_{m}}^{k_{M}}dk\,k^{2}P(k)\int_{\mathcal{N}_{i}}^{\mathcal{N}_{0}}d\mathcal{N}_{1}\,e^{-\mathcal{N}_{1}}\,\phi(\mathcal{N}_{1})\,b_{s}\,D(\mathcal{N}_{1})\,j_{l}(k\chi_{1})\,\times
\nonumber \\
& & \qquad \times
\int_{\mathcal{N}_{i}}^{\mathcal{N}_{0}}d\mathcal{N}_{2}\,e^{-\mathcal{N}_{2}}\,\phi(\mathcal{N}_{2})b_{s}\,D(\mathcal{N}_{2})j_{l}(k\chi_{2})\,.
\end{eqnarray}
In what follows, we will proceed along the same lines discussed in~\cite{01790} to calculate all the quantities that are relevant for our discussion. The required approximations are presented in the following section.

\section{\label{sectionapprox}The quasistatic and Limber approximations} 

In order to discuss the evolution of matter perturbations relevant to large-scale structure, we are particularly interested in the modes deep inside the Hubble radius (namely, those for which $k^{2}/a^{2}\gg H^{2}$). On such scales, which are, indeed, the relevant ones for ISW data, we can conveniently use the quasistatic approximation, under which the dominant contributions come from terms including $k^{2}/a^{2}$ or $\delta$~\cite{4242}. After defining $\eta\equiv\Psi/\Phi$, we can write 
\begin{equation}
\psi_{{\rm ISW}}  =  \Phi+\Psi=\Psi\left(\frac{1+\eta}{\eta}\right)\,.
\end{equation}
Then, the Poisson equation under the quasistatic approximation can be rewritten as
\begin{equation}
-\frac{k^{2}}{a^{2}}\,\Psi  =  4\pi G_{{\rm eff}}\rho_{m}\delta=4\pi G_{N}\frac{G_{{\rm eff}}}{G_{N}}\rho_{m}\delta 
 =  \frac{3}{2}\,8\pi G_{N}\frac{G_{{\rm eff}}}{G_{N}}\frac{\rho_{m}}{3}\delta =\frac{3}{2}\,\frac{G_{{\rm eff}}}{G_{N}}\frac{\rho_{m}}{3\Mpl^{2}}\delta\,. \\
 \end{equation}
We can use the last equality to find $\Psi$ in terms of $\delta$, and then
 \begin{eqnarray}
\psi_{{\rm ISW}} & = & \Psi\left(\frac{1+\eta}{\eta}\right)=  -3\,\frac{G_{{\rm eff}}}{G_{N}}\left(\frac{1+\eta}{2\eta}\right)\frac{\rho_{m0}}{3\Mpl^{2}H_{0}^{2}k^{2}}\,H_{0}^{2}\,(1+z)\,\delta\nonumber\\
 &  = & -\frac{3H_{0}^{2}\,\Omega_{m0}}{k^{2}}\,(1+z)\,\Sigma\,\frac{D(z)}{D(z=0)}\,\delta(\bm{k},z=0)\,, \label{eq:psiisw}
\end{eqnarray}
where in the last step we have introduced
\begin{equation}
\Sigma\equiv\frac{G_{{\rm eff}}}{G_{N}}\left(\frac{1+\eta}{2\eta}\right).
\end{equation}
In our model, under the quasistatic approximation $\eta=1$, $\Sigma$ has a direct physical meaning, and can be rewritten as
\begin{eqnarray}
\Sigma & \approx & \frac{G_{{\rm eff}}}{G_{N}}\nonumber \\
 & \approx & \frac{[2q\,(4\,s+3)-3]\,r_{2}+3-2q\left(6\,s+5\right)}{8\,r_{2}^{2}q^{2}s^{2}+[8\,q^{2}s+2q\,(4\,s+3)-3]\,r_{2}+3-2\left(6\,s+5\right)q}\,.\label{eq:sigma}
\end{eqnarray}
We note that at early times, i.e. $r_{2} \to 0$, we have
\begin{equation}
\lim_{r_{2}\to0}\Sigma\to1\,,
\end{equation}
and we recover GR in the limit, i.e. $G_{\rm eff}=G_{N}$ at high redshifts. The very same limit is recovered for small $s \approx 0$, as
\begin{equation}
\lim_{s\to0}\Sigma\to1\,.
\end{equation}
Another interesting limit is the late time one, when the solution approaches dS solution, i.e. $r_{2} \to 1$. In this limit we have
\begin{equation}
\Sigma_{{\rm dS}}\equiv\lim_{r_{2}\to1}\Sigma=\frac{1}{1-2qs}\,,
\end{equation}
which is positive in the region of the parameter space that we are considering, in view of~(\ref{eq:cond1})). Models for which $q\approx1/(2s)$, will be, in general, in tension with observations, as this would result in $\Sigma \gg 1$ at the present epoch. In fact, for the covariant cubic Galileon we have $s=1/2q$, which would result in arbitrarily large $G_{\rm eff}/G_N$ on the tracker solution: it is well known that this last solution is ruled out by both background~\cite{Nesseris:2010pc,CosmTsuji} and ISW~\cite{Renk:2017rzu} constraints. As we will discuss later on, for our extended Galileon model the viable parameter space corresponds to the case where $2qs$ will be smaller than unity.

We now report some relations that can be useful for computations. First,
\begin{equation}
\frac{\partial}{\partial\eta}\psi_{{\rm ISW}}  =   -\frac{3}{k^{2}D_{0}}\,H_{0}^{2}\,\Omega_{m0}\,\delta(\bm{k},0)\,\frac{\partial}{\partial\eta}[(1+z)\,\Sigma(z)\,D(z)]\,.
\end{equation}
Then, this allows to write $Z_{\rm ISW}$ as 
\begin{equation}
Z_{\rm ISW} = -\frac{3}{k^{2}}\,H_{0}^{2}\,\Omega_{m0}\,\frac{\partial}{\partial\mathcal{N}}[e^{-\mathcal{N}}\,\Sigma(\mathcal{N})\,D(\mathcal{N})]\,.
\end{equation}

On the other hand, after neglecting the presence of radiation at late times, the growth function satisfies the following perturbation equation
\begin{equation}
\ddot{D}+2H\,\dot{D}-4\pi G_{{\rm eff}}\,\rho_{m}\,D=0\,. \label{eq:deltadot}
\end{equation}
By using the e-fold variable $\mathcal{N}$, together with~(\ref{eq:Heq}) and~(\ref{eq:Hdoteq}), we can rewrite~(\ref{eq:deltadot}) as
\begin{equation}
D''+\frac{1+(4\,s+3)\,r_{2}}{2\,r_{2}s+2}\,D'-\frac{3}{2}\,\Sigma\,(1-r_{2})\,D=0\,.\label{eq:dynHighK}
\end{equation}
In the deep matter era, as $r_2\to0$ and $\Omega_{\mathrm{r}}\approx0$, we get the approximated dynamics
\begin{equation}
D''+\frac{1}{2}\,D'-\frac{3}{2}\,D\approx0\,,
\end{equation}
which has two independent solutions: $\delta_{{\rm decaying}}\propto e^{-3\mathcal{N}/2}$,
and $\delta_{{\rm growing}}\propto e^{\mathcal{N}}$ as in GR. This is consistent, since as we have already noticed $\Sigma\to1$ at early times.

An additional approximation allows to rewrite the relevant quantities in a more convenient form for large $l$, which is what we are interested in. This is based on Limber expansion, which makes use of a series expansion in powers of $(l + 1/2) ^{-1}$. Truncating the series at first order (Limber approximation) is accurate for the calculation of correlations for fields that are slowly varying in redshift. In our case, the selection of the window function that we discussed above fulfils this requirement. We can then use the result
\begin{equation}
\int k^{2}dk\,j_{l}(k\chi_{1})j_{l}(k\chi_{2})F(k)\approx\frac{\pi}{2}\,\frac{\delta(\chi_{1}-\chi_{2})}{\chi_{1}^{2}}\,F(l_{12}/\chi_{1})\,,
\quad
\mathrm{where}
\quad
l_{12}\equiv l+\frac{1}{2}
\,,
\end{equation}
and $j _{l}$ are the spherical Bessel functions of rank $l$. By using this approximation we can rewrite Eq.~(\ref{eq:ClGI}) as
\begin{eqnarray}
C_{l}^{{\rm GI},\alpha}
& \approx & 
6\pi^{2}\,\bar{\delta}_{H}^{2}\,\Omega_{m0}\lim_{\mathcal{N}\to0}\int_{\mathcal{N}_{i}}^{\mathcal{N}}d\mathcal{N}\left(-\frac{e^{-\mathcal{N}}}{l_{12}^{2}}\right)\,\frac{H}{H_{0}}\,\{[\Sigma'-\Sigma]\,D+\Sigma\,D'\}\phi^{\alpha}\,D\,b_{s}^{\alpha}\,\times\nonumber \\
&  & {}\times[T_{m}(l_{12}H_{0}/\bar{\chi})]^{2}\left(\frac{l_{12}}{\bar{\chi}}\right)^{n_{s}}=\,C_{l}^{\mathrm{GI},\alpha \mathrm{L}} \label{eq:CGI}
\,,
\end{eqnarray}
where we have substituted $P(k)$ using~(\ref{eq:powerspectrum}), and the upper label `L' stays for Limber. Moreover, we have explicitly indicated the dependence from the dataset, $\alpha$, for instance in $\phi^{\alpha}$ and $b_{s}^{\alpha}$, and we have introduced $\bar{\delta}_{H}\equiv\delta_{H}/D_{0}$. In the same way we also obtain
\begin{eqnarray}
C_{l}^{{\rm GG},\alpha \mathrm{L}} & = &
2\pi^{2}\bar{\delta}_{H}^{2}(b_{s}^{\alpha})^{2}\int_{\mathcal{N}_{i}}^{\mathcal{N}_{0}}d\mathcal{N}\,\biggl(\frac{e^{-\mathcal{N}}}{l_{12}^{2}}\biggr)\,\frac{H}{H_{0}}\,[\phi^{\alpha}]^{2}\,D^{2}\,[T_{m}(l_{12}H_{0}/\bar{\chi})]^{2}\left(\frac{l_{12}}{\bar{\chi}}\right)^{n_{s}+2} \,.
\end{eqnarray}

The integral in~(\ref{eq:CGI}) can be numerically calculated by solving the following system of coupled ordinary differential equations
\begin{eqnarray}
\frac{dD}{d\mathcal{N}} & \equiv & \pi_{D}\,,\\
\frac{d\pi_{D}}{d\mathcal{N}} & = & -\frac{1+(4\,s+3)\,r_{2}}{2\,r_{2}s+2}\,\pi_{D}+\frac{3}{2}\,\Sigma\,(1-r_{2})\,D\,,\\
\frac{dr_{2}}{d\mathcal{N}} & = & \frac{3\left(s+1\right)r_{2}\left(1-r_{2}\right)}{1+s\,r_{2}}\,,\\
\frac{d\bar{\chi}}{d\mathcal{N}} & = & -\left(\frac{1-r_{2}}{1-r_{2,0}}\right)^{1/3}\left(\frac{r_{2}}{r_{2,0}}\right)^{1/(6+6s)},\\
\frac{dZ_{l}^{G,\alpha}}{d\mathcal{N}} & = & \frac{1}{l_{12}^{2}}\left(\frac{1-r_{2}}{1-r_{2,0}}\right)^{1/3}\left(\frac{r_{2,0}}{r_{2}}\right)^{5/(6+6s)}[T_{m}(l_{12}H_{0}/\bar{\chi})]^{2}\left(\frac{l_{12}}{\bar{\chi}}\right)^{n_{s}+2}[\phi^{\alpha}]^{2}\,D^{2}\,,\\
\frac{dZ_{l}^{I,\alpha}}{d\mathcal{N}} & = & -\frac{1}{l_{12}^{2}}\left(\frac{1-r_{2}}{1-r_{2,0}}\right)^{1/3}\left(\frac{r_{2,0}}{r_{2}}\right)^{5/(6+6s)}\,\times
\nonumber\\
& & \qquad \times [T_{m}(l_{12}H_{0}/\bar{\chi})]^{2}\left(\frac{l_{12}}{\bar{\chi}}\right)^{n_{s}}\{[\Sigma'-\Sigma]\,D+\Sigma\,\pi_{D}\}\,\phi^{\alpha}\,D\,,
\end{eqnarray}
where
\[
\Sigma'=\frac{3\left(s+1\right)sq^{2}\left(r_{2}-1\right)r_{2}\left[s\left(sq+\frac{3}{4}\,q-\frac{3}{8}\right)r_{2}^{2}-3\,\left(sq+\frac{5}{6}\,q-\frac{1}{4}\right)sr_{2}-\frac{3}{2}\,sq-\frac{5}{4}\,q+\frac{3}{8}\right]}{\left(r_{2}s+1\right)\left\{ r_{2}^{2}q^{2}s^{2}+\left[(q^{2}+q)s+\frac{3}{4}\,q-\frac{3}{8}\right]r_{2}-\frac{3}{2}\,sq-\frac{5}{4}\,q+\frac{3}{8}\right\} ^{2}}\,,
\]
and $Z_{l}^{G,\alpha}$ and $Z_{l}^{I,\alpha}$ are defined as follows
\begin{eqnarray}
C_{l}^{{\rm GG},\alpha L} & = & 2\pi^{2}\bar{\delta}_{H}^{2}\,(b_{s}^{\alpha})^{2}\,Z_{l}^{G,\alpha}\,,\\
C_{l}^{{\rm GI},\alpha L} & = & 6\pi^{2}\bar{\delta}_{H}^{2}b_{s}^{\alpha}\,\Omega_{m0}\,Z_{l}^{I,\alpha}\,.
\end{eqnarray}
In terms of the above quantities the initial conditions are
\begin{eqnarray}
D(\mathcal{N}_{i}) & = & e^{\mathcal{N}_{i}}\,,\\
\pi_{D}(\mathcal{N}_{i}) & = & D'(\mathcal{N}_{i})=e^{\mathcal{N}_{i}}\,,\\
r_{2}(\mathcal{N}_{i}) & = & r_{2i}\,,\\
\bar{\chi}(\mathcal{N}_{i}) & = & \bar{\chi}_{i}\,,\\
Z_{l}^{G,\alpha}(\mathcal{N}_{i}) & = & 0\,,\\
Z_{l}^{I,\alpha}(\mathcal{N}_{i}) & = & 0\,.
\end{eqnarray}
This choice of initial conditions looks natural once we remember that at high redshift (i.e.\ in the deep matter era, namely for $\mathcal{N}_i\approx-6$) the theory reduces to GR.

\section{\label{sectionresults}Results and general bounds on the parameters of the theory}

\begin{figure}[t]
	\centering
	\subfloat[Sign of the ISW-Galaxy cross correlation for each point]{\includegraphics[width=10.5truecm]{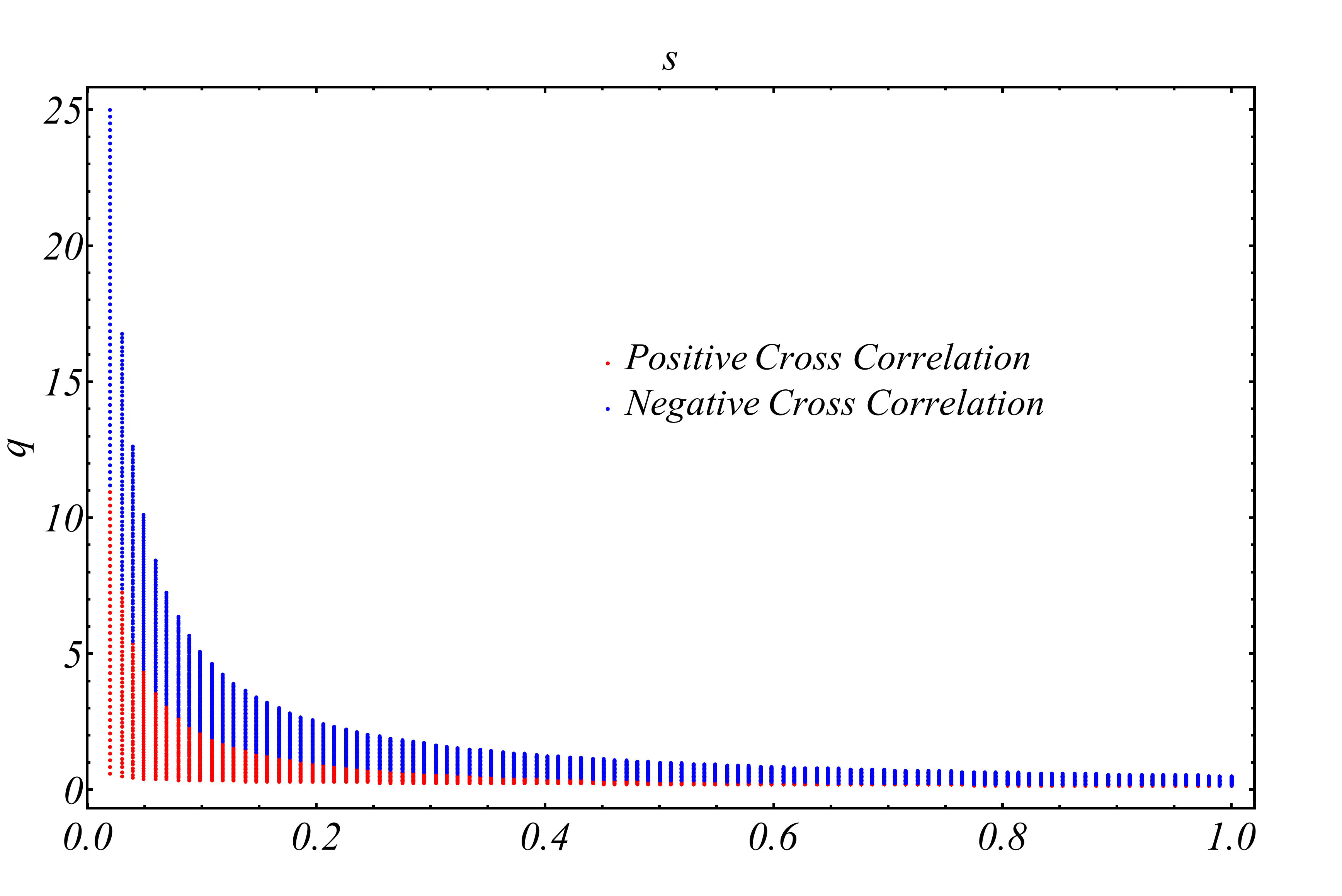}}\qquad
	\subfloat[Values of $\Sigma \equiv G_{\rm eff}/G_{N}$ for different $q$ and $s$]{\includegraphics[width=10.5truecm]{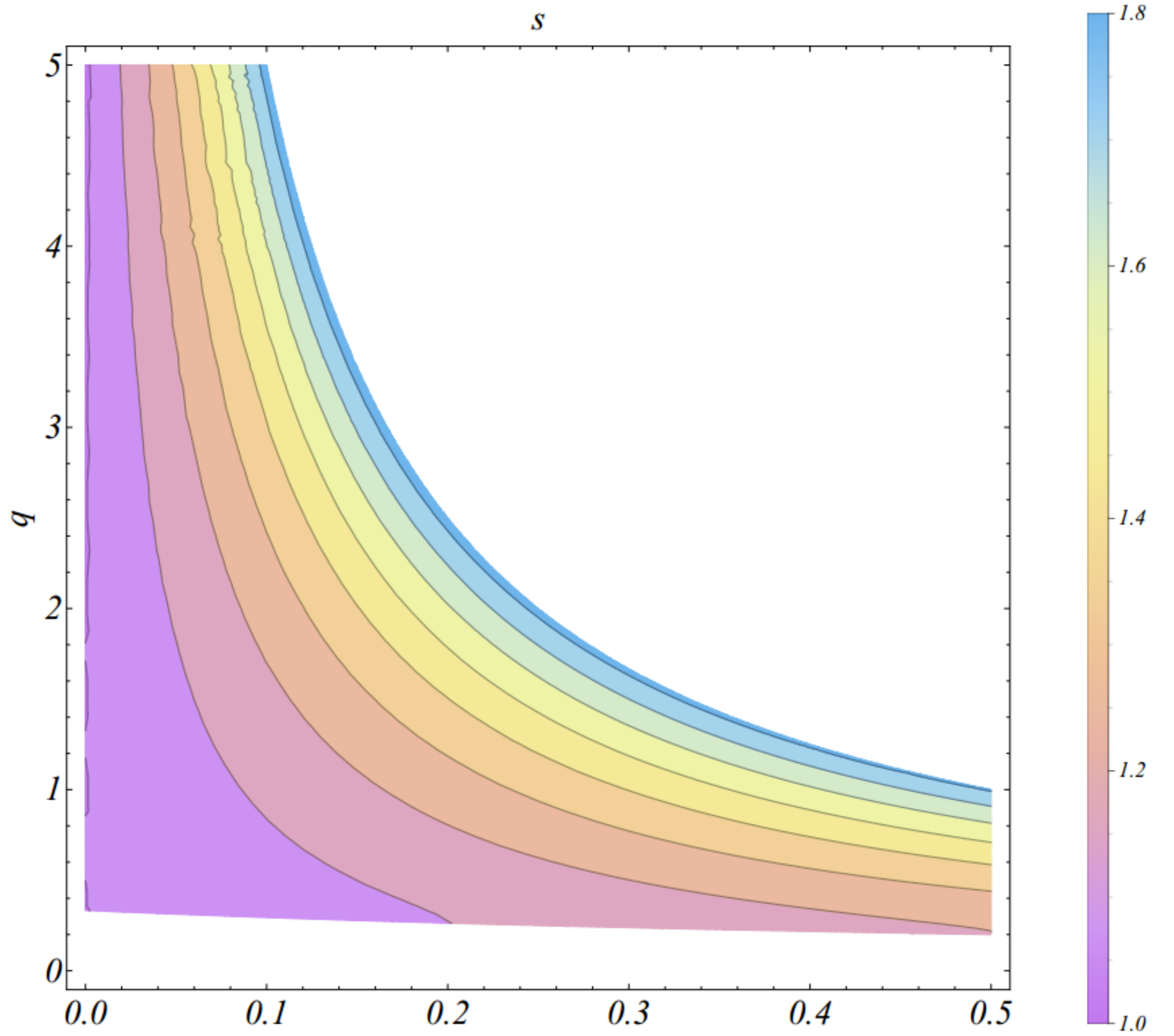}} \\	
	\caption{Results of the analysis for the whole available parameter space}
	\label{fig:wholespace}
\end{figure}
In this section we present the results of our numerical computations for the study of the dependence on the model parameters $q$, $s$, of the chi-squared for the ISW-Galaxy cross correlation, $\chi^2_{\rm ISW}$. In particular, we assume as free parameters only the model parameters $q$ and $s$, whereas we keep fixed the other cosmological parameters, e.g., $\Omega_{\mathrm{m}0}$ and $h$, to the best-fit values provided by Planck \cite{Planck}. A general parameter space analysis will be discussed later, but, first, we would like to focus on the behaviour of $\chi_{\rm ISW}^2$ as a function of $q$ and $s$, in order to have an idea of how much the two-dimensional parameter space can be restricted due to the presence of negative cross-correlation.

\begin{figure}[ht]
	\centering
    \includegraphics[width=13truecm]{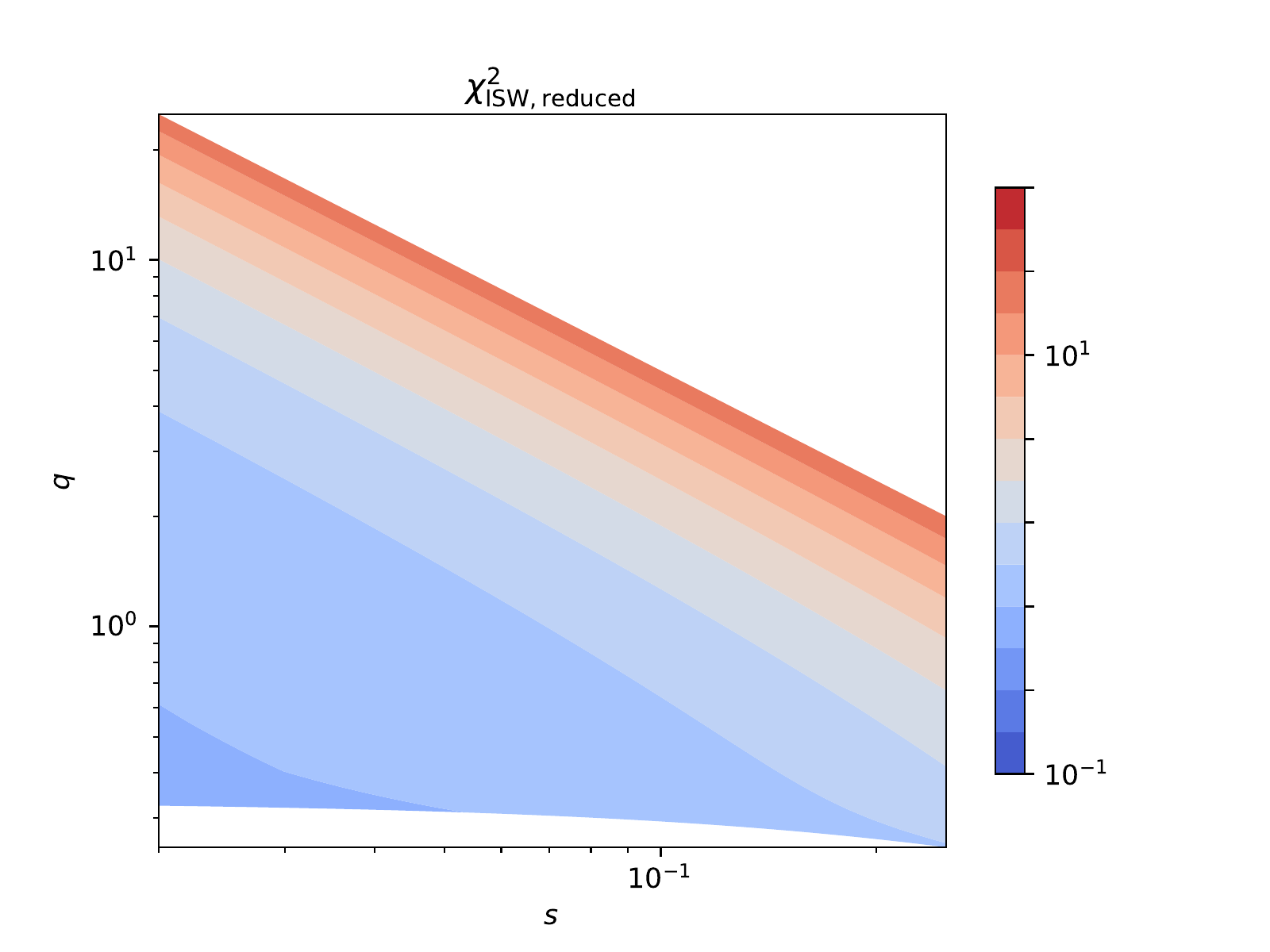}
	\caption{Reduced chi-square for the ISW-Galaxy cross correlations data as a function of $s$ and $q$}
	\label{fig:lowpspace}
\end{figure}

To sample the parameter space that we discussed in section~\ref{stability conditions}, we start by selecting one hundred uniformly distributed values of the variable $s$ (inside the allowed ``stability'' interval): then, for each of these values, we consider one hundred values of the variable $q$, again uniformly distributed in the allowed interval. For each of these ten thousands points we solve the ODEs system equivalent to the integral~(\ref{eq:CGI}) in order to obtain the sign of the ISW-Galaxy cross correlation. The results of this first sampling are shown in Fig.~\ref{fig:wholespace}: we see that for the majority of the points (panel (a)) the cross correlation is negative. However, there is still a non-negligible region of positive cross correlation located in the bottom left corner of the plot, i.e., for low values of $s$ and $q$. In the same figure (panel (b)) we also show how the value of $\Sigma \equiv G_{\rm eff}/G_{N}$, given by~(\ref{eq:sigma}), varies in the parameter space: notice that it is always larger than one, getting closer to the GR value in the positive cross correlation region. On the contrary, large values of $\Sigma$ tend to give anti-cross correlation. We conducted a more refined analysis by sampling another ten thousands points in the more favorable region only: the results are shown in Fig.~\ref{fig:lowpspace}, where we plot the reduced $\chi^2_{\rm ISW}$. We have considered twenty-six data points in total, coming from, both, the 2d-mass and SDSS galaxy cluster observations~\cite{4380}. In particular, for each point of the chosen grid inside the selected $(s,q)$ region we have computed the observable
\begin{equation}
w^{\alpha}(\theta)=\frac{T_{\rm CMB}}{4\pi}\sum_{l}(2l+1)\,C_{l}^{{\rm GI},\alpha}\,P_{l}(\cos\theta)\,,
\end{equation}
where $P_{l}(\cos\theta)$ are the Legendre polynomials, $T_{\rm CMB}=2.7255\,  K$ (which is the value given by the latest observations \cite{Planck}), and $\theta$ represents the angle that parametrizes deviations from the center of the galaxy data set under consideration ($\alpha$ labels the specific experiment, i.e., 2d-mass and SDSS). We compute the sum by considering $l$ ranging from $2$ to $150$, and by making use of the above mentioned Limber's approximation, i.e., we actually use $C_{l}^{\mathrm{GI},\alpha \mathrm{L}}$ in place of $C_{l}^{\mathrm{GI},\alpha}$. In our numerical/theoretical results comparison with the actual data points of 2d-mass and SDSS from~\cite{4380}, we have considered the errors given by the jackknife estimation method. The overall reduced chi-squared for each point is shown in Fig.~\ref{fig:lowpspace}. Notice that the chi-squared becomes lower and lower for values inside the positive cross correlation region closer to the GR limit.

We now move to a more detailed discussion of constraints on the model parameters, following a treatment similar to the one that can be found in~\cite{ProcaTheories} to perform a Markov Chain Monte Carlo (MCMC) sampling of the total chi-square for this model. The part of the $\chi^2$ dependent on the pure-background constraints (i.e., not considering perturbation constraints, such as redshift space distortion (RSD) and ISW) is equal to the value found in~\cite{ProcaTheories}, because it is known that the theories (on the tracker solution) have a degenerate background dynamics. However, in our model, for the perturbations constraints, namely the RSD-data (growth of perturbations) and ISW-Galaxy cross-correlation data, we obtain, in general, different results. In fact, in our theory $G_{\rm eff}/G_N$ will differ from the value obtained in generalized Proca theories. 

\begin{figure}[ht]
	\centering
	\includegraphics[width=15.5truecm]{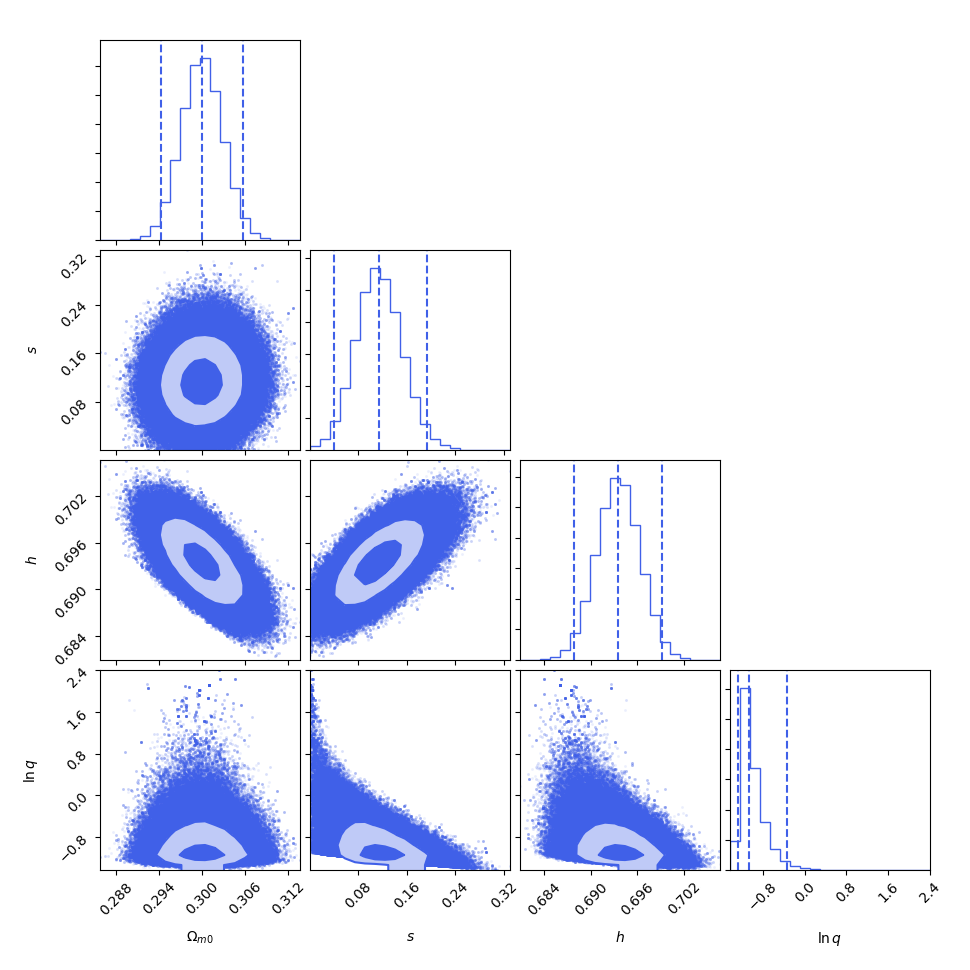}
	\caption{MCMC sampling for the parameters on the model.}
    \label{fig:montec}
\end{figure}

We use EMCEE~\cite{emcee} in order to perform MCMC-sampling. For this scope, we have set flat prior on the value of $\ln(q)$, and excluded the parameter space for which instabilities (or strong coupling) occur.  The results are in Fig.~\ref{fig:montec}

In particular, the results can be summarized as.
\begin{eqnarray}
\Omega_{m0} &=& 0.300 \pm 0.006\,,\\
s &=& 0.11 \pm 0.08\,,\\
h &=& 0.693 \pm 0.006\,,\\
q &=& 0.37 ^{+ 0.70}_{- 0.08}\,.
\end{eqnarray}
In this case, once more, the tension in the data for the value of $h$, tends to set a value of $s$ different from zero. Moreover, in this region of the parameter space we find that the best fit value is given by
\begin{equation}
\Omega_{m0} \approx 0.30,\quad s \approx 0.12,\quad h \approx 0.69,\quad q \approx 0.29\, .
\end{equation}
We can say that the effect of the ISW-Galaxy cross-correlation data is to shift the best fit for the value of $s$ to lower values compared to the background best fit value, which is similar to what is found in the Proca theories~\cite{ProcaTheories}.

\section{Summary and Conclusions} \label{sectionconclusion}
In our work, we consider a particular class of generalized covariant Galileon models~\cite{3878}, for which we compute the dynamics of both the background and the scalar perturbations. Compared to $\Lambda$-CDM this model introduces two extra parameters, $s$ and $q$ (the latter affects only the dynamics of the perturbations). Then, we determine the parameter subspace, where the model is stable and, at the same time, gives positive ISW-Galaxy cross correlation. Our aim was to evaluate the ISW-Galaxy cross correlation, and check whether this model is excluded by such an observable. Contrary to what is sometimes given for granted by intuitively extending the results obtained in other models, we find that there is, indeed, a parameter subspace for which the ISW-Galaxy cross correlation turns out to be positive. This result clearly shows that in some appropriate version (e.g., our generalized version) the cubic Galileon model can be considered as a physically viable model. This is a concrete counterexample to what is commonly assumed for $G_3$ Horndeski theories. One can see from Fig.~\ref{fig:wholespace} that the effective gravitational constant $G_{\rm eff}$ gets closer and closer to the GR value for lower values of $q$: this region corresponds to the positive cross correlation region. In fact, as can be seen from Fig.~\ref{fig:lowpspace}, in this very region also the value of the $\chi^{2}_{\rm ISW, reduced}$ becomes reasonable, while it grows substantially moving to higher values of $q$ and $s$. The reason for the anti-cross correlations, which occurs for larger values of $q$ and $s$, can be traced to the increase of the value of the effective gravitational constant, i.e., $G_{\rm eff}$, in this region of the parameter space.

After performing a general parameter space analysis, we find that non-zero values of $s$ are favoured by data sets. This is consistent with what is found in the context of Proca theories~\cite{ProcaTheories}. In fact, a positive value of $s$ is capable of reducing the tension on the background value of today's Hubble parameter $H_0$. The ISW constraints contribute to a shift of the preferred best fit value for $s$ towards lower (but still positive) values.

We hope that this study will give motivation to the use of general cubic Horndeski $G_3$-terms for theories that require these kind of Lagrangians in order to implement a viable Vainshtein mechanism, e.g.~\cite{minquasi}, to pass Solar system constraints.

\section*{Acknowledgements}

A.D.F.\ was supported by JSPS KAKENHI Grant No.\ 16K05348. S.A. would like to gratefully acknowledge hospitality by the Department of Physics of Kyoto University, and partial support by the Italian Institute of Nuclear Physics (INFN). F.G. wishes to thank the Department of Physics of Kyoto University and the Yukawa Institute for Theoretical Physics for the warm hospitality.

\end{document}